# Evolution of medium-range order and its correlation with magnetic nanodomains in Fe-Dy-B-Nb bulk metallic glasses

Jiacheng Ge,[a] Yao Gu,[a] Zhongzhen Yao,[a] Sinan Liu,[a] Huiqiang Ying,[a] Chenyu Lu,[b] Zhenduo Wu,[b,c] Yang Ren,[b] Jun-ichi Suzuki,[d] Zhenhua Xie,[e] Yubin Ke,[e,f,*] He Zhu,[a] Song Tang,[a] Xun-Li Wang,[b,g] and Si Lan[a,b,*]

[a]Herbert Gleiter Institute of Nanoscience, School of Materials Science and Engineering, Nanjing University of Science and Technology, 200 Xiaolingwei, Nanjing 210094, China
[b]Department of Physics, City University of Hong Kong, Hong Kong SAR, China
[c]City University of Hong Kong (Dongguan), Dongguan 523000, China
[d]Japan Proton Accelerator Research Complex, Japan Atomic Energy Agency, Tokai, Japan
[e]Spallation Neutron Source Science Center, Dongguan 523803, China
[f]Guangdong-Hong Kong-Macao Joint Laboratory for Neutron Scattering Science and Technology, 1 Zhongziyuan Road, Dalang, Dongguan 523803, China
[g]Center for Neutron Scattering, City University of Hong Kong Shenzhen Research Institute, Shenzhen 518057, China

*Corresponding author(s): lansi@njust.edu.cn; keyb@ihep.ac.cn

**Abstract**

Fe-based metallic glasses are promising functional materials for advanced magnetism and sensor fields. Tailoring magnetic performance in amorphous materials requires a thorough knowledge of the correlation between structural disorder and magnetic order, which remains ambiguous. Two practical difficulties remain: the first is directly observing subtle magnetic structural changes on multiple scales, and the second is precisely regulating the various amorphous states. Here we propose a novel approach to tailor the amorphous structure through the liquid-liquid phase transition. In-situ synchrotron diffraction has unraveled a medium-range ordering process dominated by edge-sharing cluster connectivity during the liquid-liquid phase transition. Moreover, nanodomains with topological order have been found to exist in composition with liquid-liquid phase transition, manifesting as hexagonal patterns in small-angle neutron scattering profiles. The liquid-liquid phase transition can induce the nanodomains to be more locally ordered, generating stronger exchange interactions due to the reduced Fe–Fe bond and the enhanced structural order, leading to the increment of saturation magnetization. Furthermore, the increased local heterogeneity in the medium range scale enhances the magnetic anisotropy, promoting the permeability response under applied stress and leading to a better stress-impedance effect. These experimental results pave the way to tailor the magnetic structure and performance through the liquid-liquid phase transition.

1. **Introduction**

Fe-based metallic glasses (MGs) are promising functional materials because of their attractive soft magnetic properties. The advantage of ultralow coercive force, high permeability, and low core loss are promising for low-energy-consumption devices [1, 2]. Also, the giant stress-impedance and giant magneto-impedance lead to excellent outfield response and magnetic functional performance in soft magnetic MGs, which can thus be used as the core of advanced magnetic sensors [3]. Numerous studies have reported the arrangement and movement of magnetic domains directly determined the magnetic performance of MGs [4-6]. The conventional strategy to tailor magnetic domains and thereby improve the magnetic performance focuses on the methods of alloying [7, 8], relaxation [9], and crystallization [10]. Still, the correlation between magnetic domains and the structure of MGs is not well established. The structure of MGs is locally ordered over the short range (2–5 Å) and medium range (5–20 Å) but disordered over the long range. Medium-range order (MRO) describes the packing of clusters with short-range order (SRO) [11, 12]. Numerous studies report that MRO plays a decisive role in tuning the properties of MGs[13], including mechanical performance [14] and catalytic performance [15]. Although it was found that MRO may correlate with magnetic order in magnetic MGs [16, 17], any such correlation remains ambiguous. Practical difficulties arise in two forms: The first is the missing experimental model system which can be precisely regulated among different amorphous states. The second is the difficulty to directly observe subtle variations in magnetic structure on a nanoscale. Traditional characterization techniques, such as magnetic force microscopy, only capture magnetic structures on a micron scale [18-20].

Recently, a series of MGs (Pd, Mg, and Zr based) [21, 22] with a liquid-liquid phase transition (LLPT) has been reported. The LLPT is a universal phenomenon that involves the cooperative arrangement of bond-orientation orders and locally favored structures upon cooling or heating. Our previous work revealed the LLPT in Fe-Y-B-Nb MGs in the anomalous exothermic region of differential scanning calorimetry (DSC) curves [23]. The results show that the structural evolution of the MRO is prominent in the atomic rearrangement. This local atomic rearrangement over the short-to-medium range can reorient the magnetic moments, thereby redistributing the magnetic domains through spin-orbit coupling. Therefore, LLPT offers an appropriate microstructure manipulation approach to tailor the local orders without destroying the amorphous structure in Fe-based MGs. The LLPT-induced evolution of MRO can further alter the magnetic domain in amorphous materials.

This work uses *in-situ* synchrotron high-energy x-ray diffraction (HE-XRD) and small-angle neutron scattering (SANS) to investigate the evolution of the atomic structure and magnetic nanodomains under magnetic fields in the Fe-Dy-B-Nb bulk MGs (BMGs) with LLPT. The $(Fe_{0.72}Dy_{0.05}B_{0.24})_{97}Nb_4$ MG has an anomalous exothermic peak (AEP) in the supercooled-liquid region. Locally ordered regions have been revealed at the temperature of AEP, then reenter the disordered state at higher temperatures. A real-space analysis reveals that enhancing the edge-sharing-cluster connectivity causes medium-range ordering during the LLPT. Moreover, two-dimensional (2D) SANS patterns find hexagon-like topological magnetic scattering signals, which are indicative of hidden nanodomains with topological structure. The magnetic correlation length of the hidden topological nanodomains undergoes the same evolution as the edge-sharing percentage in all four cluster-connection modes. The magnetic performance, including saturation magnetization and stress-impedance, is enhanced in post-LLPT samples. The experimental results reported herein pave the way for building a relationship among atomic, magnetic structures and the magnetic properties of Fe-based MGs.

## 2. Experiment

*2.1 Sample preparation*

Elemental Fe (99.9%), B (99.5%), and Nb (99.9%) were carefully weighed in the proper mass ratio and melted by induction to obtain the ingots of $Fe_{72}B_{24}Nb_4$. Dy (99.9%) was then weighed and added to the ingots by arc melting. Before suction casting, the ingot surface oxide layer was removed with a grinder to prevent heterogeneous nucleation. The result of suction casting was 2-mm-diameter, 30-mm-long glassy rods of $(Fe_{0.72}Dy_{0.05}B_{0.24})_{97}Nb_4$ and $(Fe_{0.72}Dy_{0.03}B_{0.24})_{97}Nb_4$, which are abbreviated Dy5 and Dy3, respectively. Ribbons were prepared by melt spinning.

*2.2 Characterization and performance tests*

The specific heat capacity experiment applied DSC (DSC-1 made by METTLER TOLEDO) in a high-purity $N_2$ atmosphere with a heating rate of 20 K/min.

Synchrotron high energy- X-ray diffraction (HE-XRD) for pair distribution function (PDF) analysis was conducted at beamline 11-ID-C at the Advanced Photon Source, Argonne National Laboratory. High energy X-rays with a beam size of 500μm×500μm and wavelength of 0.1173 Å were used in transmission geometry for data collection. Two-dimensional diffraction patterns were obtained by using a Perkin Elmer amorphous silicon detector with a data acquisition time of 1 s for each pattern. The time for data readout and storage was approximately 4 s. The heating rate was 20 K/min. The time resolution was about 5 s, and the temperature resolution was 1–2 K. The static structure factor, $S(Q)$ with $Q_{max} \approx 30$ Å$^{-1}$, was derived from the scattering data after applying the Laue correction and masking bad pixels, integrating images, subtracting the appropriate background, and correcting for oblique incidence, absorption, multiple scattering, fluorescence, and Compton scattering by using Fit2D and PDFgetX2. The reduced PDF $G(r)$ is obtained from the Fourier transform of $S(Q)$, $G(r) = (2/\pi) \int_0^{Q_{max}} Q(S(Q) - 1) \sin(Qr) \, dQ$, where $r$ is the distance in real space and $Q = 4\pi \sin\theta/\lambda$, where $\theta$ is half of the scattering angle between the incident beam and the scattered beam, and $\lambda$ is the x-ray wavelength.

SANS under a magnetic field was conducted at beamline 15 of J-PARC (Japan) and at the SANS of CSNS (China). The incident neutrons were unpolarized. The external magnetic field was provided by a superconducting magnet with a field direction perpendicular to the wave vector of the incoming neutron beam and in the plane of the bulk sample with a 1×1 cm$^2$ area. By using three sample-to-detector distances, this setup had an accessible $Q$ range of 0.005 to 1.5 Å$^{-1}$. Raw SANS data were corrected for background scattering and detector efficiency. The total scattering cross section d$\Sigma$/d$\Omega$ involves both nuclear and magnetic scattering. The spin-misalignment cross section d$\Sigma_M$/d$\Omega$ contains only the purely magnetic scattering due to the transversal spin components, obtained by subtracting the scattering cross section d$\Sigma$/d$\Omega$ at a saturating magnetic field of 1 T from the scattering cross section at 0.05 T [24, 25]. The autocorrelation function $C(r)$ of the spin-misalignment is obtained using $C(r) = \frac{\omega}{2\pi^2 b_m^2 \rho_a^2 r} \int_0^\infty \frac{d\Sigma_M}{d\Omega} sin(qr)qdq$, where $\omega$ = 3/2 at small applied magnetic field (demagnetized state), and $\omega$ = 4/3 for the nearly saturated texture-free ferromagnet. In the expression for $C(r)$, $b_m$ and $\rho_a$ are the atomic magnetic scattering length and the atomic density, respectively [26]. The correlation length $l_C = r$ for which $C(r) = C(0)/e$, where $C(0)$ is the value of $C(r)$ extrapolated to the origin.

The hysteresis loop was measured by using a physical property measurement system with the vibrating specimen magnetometer module (DYNACOOL-9, Quantum Design, USA). The applied magnetic field range is −10000 to 10000 Oe. The impedance $Z$ under applied tensile stress $\sigma$ was measured by using an impedance analyzer (LCR Meter IM3536 made by Hioki) at a frequency of 1 MHz and with a driving current of 100 μA. Tensile stress was applied by using an electronic universal tester (UTM4304GD made by SUNS). The measurements were made at room temperature. The stress-impedance ratio SI is defined as $\Delta Z/Z = [|Z(\sigma) - Z(0)|/Z(0)] \times 100\%$, where $Z(\sigma)$ is the impedance under the applied stress $\sigma$, and $Z(0)$ is the impedance under zero stress [27]. The effective magnetic permeability was also measured by using the IM3536 impedance analyzer. The glassy ribbons were wound into circles, and the effective magnetic permeability was obtained by using $\mu_{eff} = LL_e/(\mu_0 N^2 A_e)$, where $\mu_{eff}$ is the effective permeability, $L$ is the inductance, $L_e$ is average effective magnetic circuit length, $\mu_0$ is the vacuum permeability, $N$ is number of coil turns, and $A_e$ is effective cross-sectional area of the core [28]. $L_e = \pi(D-d)/\ln(D/d)$ and $A_e = (D-d)H/2$, where $D$ is the inner diameter of the magnetic core, $d$ is the outer diameter of the magnetic core, and $H$ is the core height.

**Results**

*3.1 Thermophysical behavior, AEP, and enhanced magnetic performance*

Fig. 1(a) shows the DSC results for Dy5 BMGs. The anomalous exothermic peak (labeled $T_{AEP}$, which is ≈914 K) in the supercooled liquid region. After annealing in the AEP region, the saturation magnetization of the sample annealed at $T_{AEP}$ is enhanced by approximately 12% compared with the as-cast state, following which the magnetization deteriorates back to the same level as the as-cast sample at a higher temperature [see Table 1 and Fig. 1(b)]. Besides, the coercive force increases slightly and remains at a very low level (≈1.25 Oe). Magnetoelastic coupling is the basis of many magnetic and stress sensors, one of which uses the stress-impedance effect[29]. Stress alters the magnetic permeability, manifested as a significant change in the impedance. The stress impedance is enhanced for the $T_{AEP}$-annealed sample and show a maximum SI of 206% at a stress of 33 MPa, which is nearly twice the improvement of the as-cast sample (Table 1). The skin effect [30] explains the stress impedance: when the ferromagnetic material is stressed, the equivalent magnetic field changes due to the magnetoelastic coupling, affecting the spontaneous magnetization of the material and changing the effective permeability. As the impedance $Z$ is in proportion to $\mu_{eff}^{0.5}$, here the change of effective magnetic permeability $\mu_{eff}$ is bigger in Dy5 after annealing [Fig. 1(d)] so that $T_{AEP}$-annealed Dy5 has a better stress impedance.

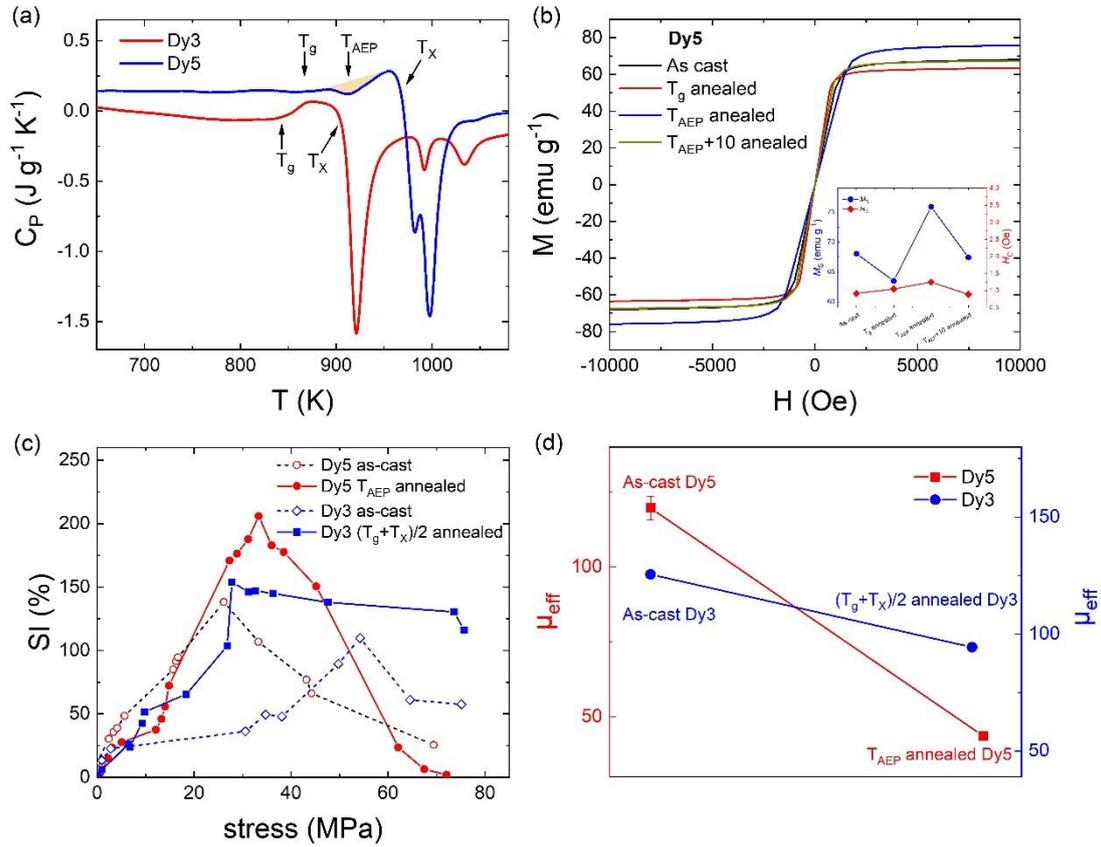

**Fig. 1.** (a) DSC curves of Fe-Nb-B-Dy bulk metallic glasses. An anomalous exothermic peak exists in the supercooled-liquid region, labeled $T_{AEP}$. (b) Hysteresis loops of Dy5 for different heat-treatment states. The illustration shows the saturation magnetization and coercivity at different heat-treatment states. (c) Stress-impedance ratio of as-cast state compared with that of annealed state in Dy5 and Dy3 MGs. (d) Evolution of the corresponding effective magnetic permeability $\mu_{eff}$ between the as-cast and annealed state.

**Table 1.** Magnetic characteristics of the as-cast state compares with those of $T_{AEP}$-annealed state for $(Fe_{0.72}Dy_{0.05}B_{0.24})_{96}Nb_4$ MG.

| State | $M_S$ (emu g$^{-1}$) | $H_C$ (Oe) | Maximum SI | $\mu_{eff}$ |
| --- | --- | --- | --- | --- |
| As cast | 67 | 0.92 | 138% | 120 |
| $T_{AEP}$ annealed | 76 | 1.25 | 206% | 44 |

*3.2 In-situ synchrotron x-ray diffraction study of liquid-liquid phase transition*

To reveal the underlying correlation between an amorphous structure and magnetic performance, in-situ synchrotron x-ray diffraction is conducted. Fig. 2(a) shows the structure factor $S(Q)$ of the Dy5 BMG upon heating. Fig. 2(c) shows the corresponding difference $S(Q)$, obtained by subtracting the diffraction pattern at 799 K (this temperature is less than $T_g$). $Q_1$ is the peak position, and the second moment of $Q_1$ is related to the peak width of the first sharp diffraction peak in $S(Q)$[23]. As shown in Figs. 3(b) and S2(a) of the Supplementary Material, an unusual change in slope occurs for the peak position and peak width at 914 K, which matches well with the AEP temperature. In addition, the peak height of the first sharp diffraction peak also rises and falls around $T_{AEP}$. The first moment of $Q_1$ is linked to the density, and the second moment of $Q_1$ and the peak height are related to the correlation length for the glasses and liquids, respectively [31, 32]. For comparison, $Q_1$ and the peak height are almost linear in temperature in the supercooled liquid region for the Dy3 BMG, which has no AEP (Fig. S1). This transition indicates that the Dy5 supercooled liquid initially becomes more ordered and then returns to a less-ordered state at a high temperature after $T_{AEP}$, indicating a reentrant behavior of the atomic structure in the supercooled liquid upon heating.

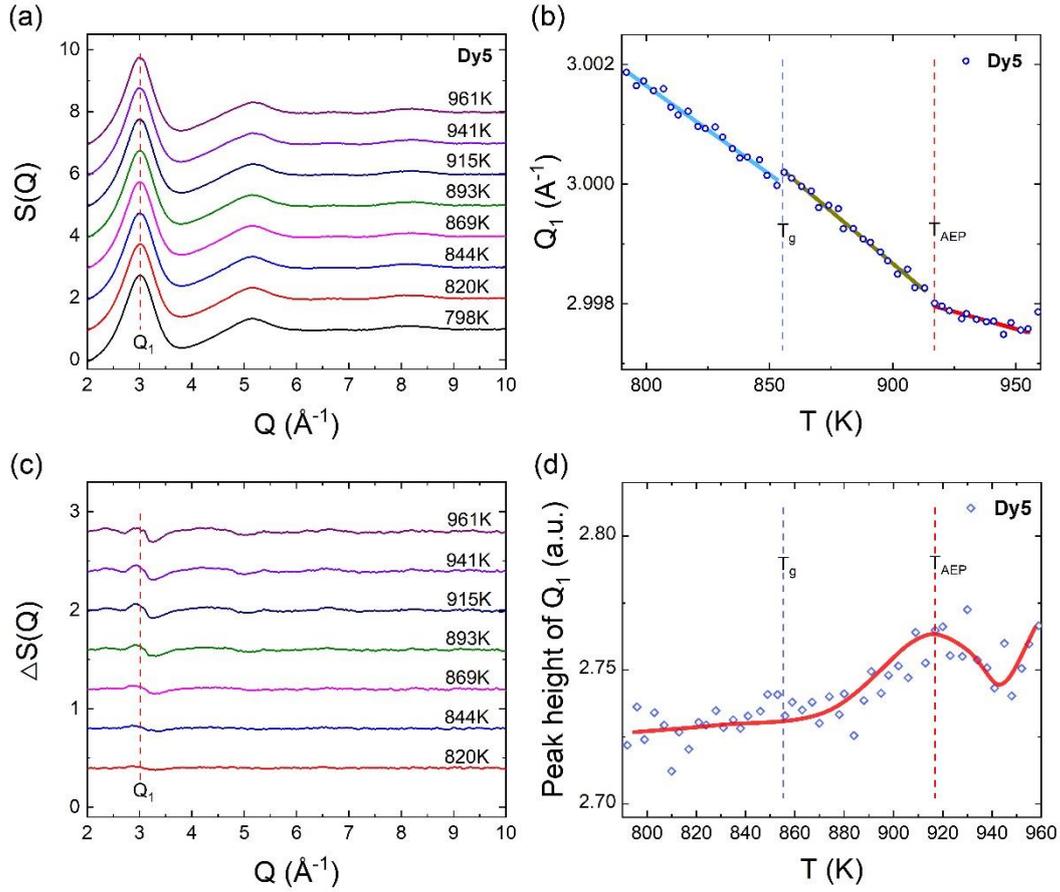

**Fig. 2.** Structure factor of Dy5 bulk metallic glasses: (a) $S(Q)$ of Dy5 BMG, (c) $\Delta S(Q)$ in Dy5 BMG, (b) first moment $Q_1$ and (d) peak height in Dy5 BMG.

Fig. 3(a) shows the reduced PDF $G(r)$ obtained by Fourier transform of $Q(S(Q) - 1)$. The first peak $R_1$ in $G(r)$ corresponds to the nearest-neighbor shell and can be used to identify the structural information of solute-centered SRO. The shoulder $R_{22}$ of the second peak indicates the packing connectivity of SRO on a medium-range scale. Fig. 3(b) shows the integrated intensity for $R_1$ [integrated region of $G(r) \geq 0$] and the integrated intensity for $R_{22}$ (integrated region ≈4.54–4.55 Å). The intensity of $R_{22}$ for Dy5 rises and falls at $T_{AEP}$ while the intensity of $R_1$ remains essentially constant, indicating that MRO plays a dominant role in this process. This behavior was confirmed in our previous study of Fe-Y-Nb-B MGs [21]. The composition with AEP shows that a cooperative rearrangement dominated by MRO plays an important role in stabilizing the supercooled liquid.

Here we do further analysis to reveal the structure evolution at medium range. Fig. 3(c) shows the radial distribution functions $g(r)$ obtained by the transform $4\pi\rho_0 G(r) + 1$, through which we can analyze the cluster connection over the medium-range scale. The neighboring SRO packing can be divided into four types: polyhedra sharing of one, two, three, or four atoms to connect with each other, denoted hereinafter as 1-, 2-, 3-, and 4-atom connections, respectively. The first three categories refer to connections that share a polyhedron vertex, edge, or face, while the last category (i.e., 4-atom) shares distorted quadrilateral or squashed tetrahedra (i.e., with 4 atoms almost in the same plane, but not necessarily forming a perfect quadrangle face) [33]. In Fig. 3(d), the 2-atom connection accounts for about 50% of the four connection modes and also rises and falls near $T_{AEP}$, whereas the 3-atom connection takes the opposite path. As a result, the SRO tends to pack in the form of edge-sharing (2-atom connection) instead of face-sharing (3-atom connection), which contributes to the anomalous change in intensity of $R_{22}$ in $G(r)$. The 3-atom connection increases sharply upon crystallization, indicating that this connection mode favors crystal formation. In the composition without AEP, all connection modes of Dy3 change little before crystallization [Fig. S3(b)].

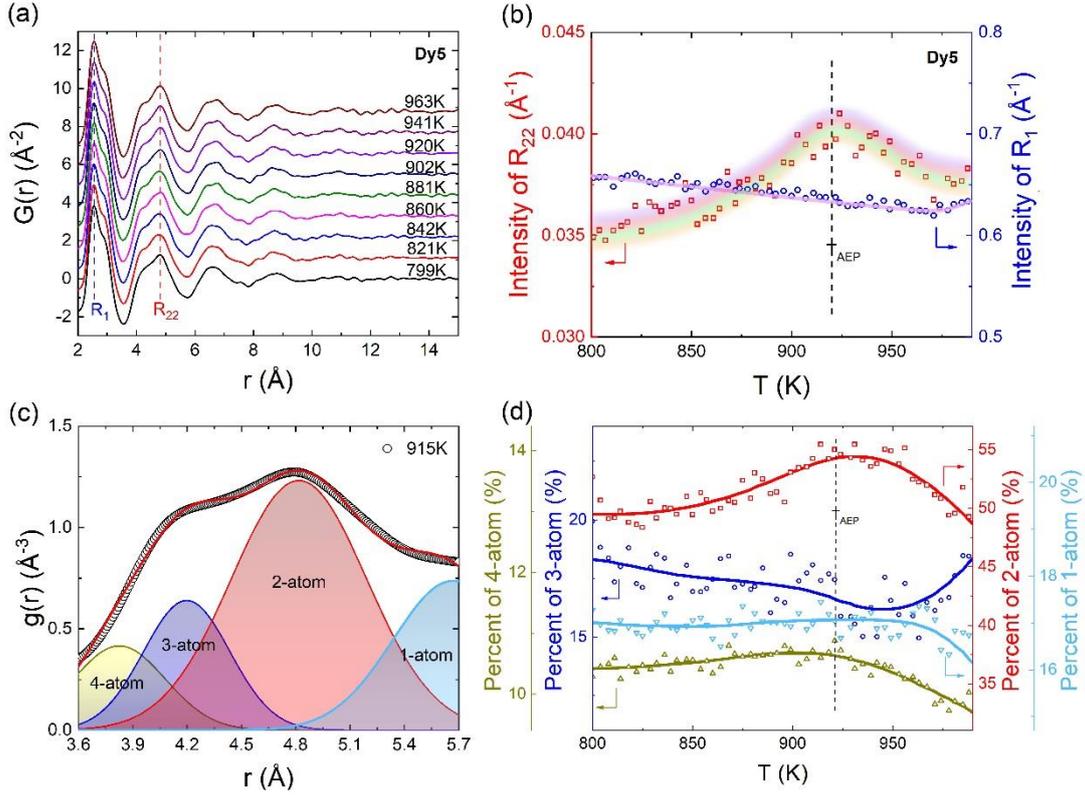

**Fig. 3.** Real-space analysis of the synchrotron x-ray diffraction results. (a) Reduced pair distribution function $G(r)$ for Dy5 BMG upon heating. (b) Intensity of first coordination shell $r_1$ (right axis) and shoulder of second coordination shell $r_{22}$ (left axis) as functions of temperature. (c) Gaussian fitting of second coordination shell in pair distribution function $g(r)$ at $T_{AEP}$. The decomposed peaks are the corresponding 1-, 2-, 3-, and 4-atom cluster connections. (d) Evolution of corresponding atomic connectivity. The percent of different connection modes as a function of temperature.

*3.3 Magnetic response to liquid-liquid phase transition at nanoscale*

Fig. 4(a) shows the two-dimensional SANS pattern of $T_{AEP}$-annealed Dy5, which is dumbbell-shaped with the maximum intensity at $\phi = 90°$. Moreover, the abnormal intensities are located at approximately $\pm 0.12$ Å$^{-1}$ in the $Q_x$ and $Q_y$ direction, which manifests as interference peaks at approximately $\pm 0.12$ Å$^{-1}$ in the one- dimensional scattering cross section $d\Sigma/d\Omega$-$Q$ plot in both Figs. 4(b) (under an unsaturated magnetic field) and 5(a) (under a saturated magnetic field). However, as displayed in Fig. 5(b), the interference peaks disappear without applying a magnetic field, reflecting the field-dependent nature of these peaks. Furthermore, the vertical $d\Sigma/d\Omega$ contains both magnetic and nuclear scattering, while the horizontal contains only nuclear scattering. Therefore, only the azimuthally averaged SANS cross section $d\Sigma/d\Omega$ in the vertical direction under 0.05 T field show the interference peaks [Fig. 5(c)], further confirming that the interference peaks originate in the nanoscale magnetic structure.

To gain more information about the magnetic structure, we apply a real-space analysis. The spin-misalignment SANS cross section $d\Sigma_M/d\Omega$ (Fig. S4) contains purely magnetic scattering due to the transversal spin components [24, 25] and is obtained by subtracting the scattering cross section $d\Sigma/d\Omega$ at 1 T saturating field from $d\Sigma/d\Omega$ measured at 0.05 T. The corresponding autocorrelation function $C(r)$ is obtained by Fourier transform of the extracted $d\Sigma_M/d\Omega$. Fig. 4(c) shows the normalized $C(r)$ of Dy5 after isothermal annealing at different temperatures. The calculated correlation length $l_C$ of $C(r)$ can serve to determine the size of the inhomogeneous magnetized regions (the calculation is given in the Experiment section). The inset of Fig. 4(c) shows how the correlation length evolves across the AEP region. The result shows that $l_C \approx 20$ Å is in the MRO. Moreover, $l_C$ rises and falls around $T_{AEP}$, which is consistent with the evolution of MRO.

The $d\Sigma/d\Omega$-$Q$ and $d\Sigma_M/d\Omega$-$Q$ curves in the low-$Q$ range are usually described by the power-law relation $K/q^n$ [34, 35], for which the fitted exponents are plotted in Fig. 4(d). The magnetic power-law exponent in $d\Sigma_M/d\Omega$ contains contributions from the magnetic anisotropy and magnetization jumps at internal interfaces [26]. The $T_{AEP}$-annealed Dy5 samples have much larger exponents $n$, indicating that the LLPT induces the enlarged inhomogeneous magnetization regions. After higher-temperature annealing ($T_{AEP}$ + 10 K), the exponent $n$ and $l_C$ return to the same level as prior to the LLPT. The temperature-dependent evolution of magnetic heterogeneity of Dy5 samples in three different states was also investigated by using SANS at CSNS (Fig. S5), which gives very consistent results.

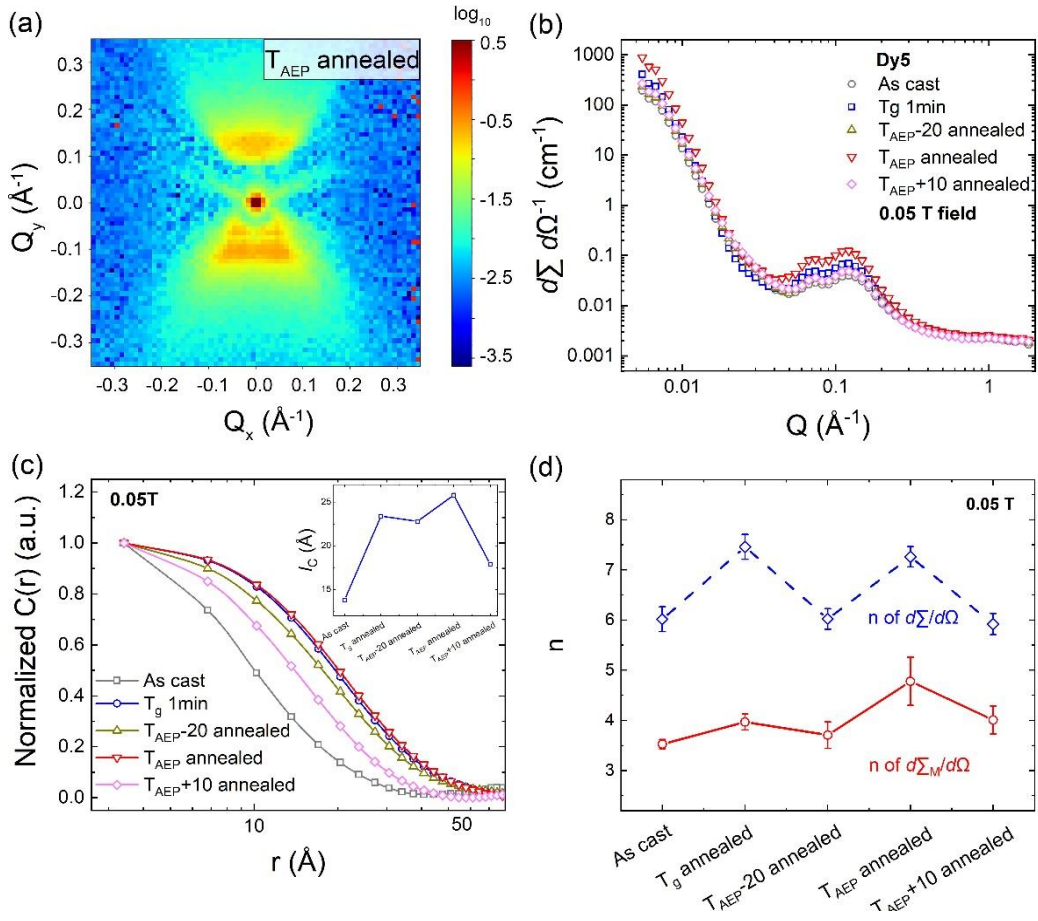

**Fig. 4** (a) 2D total scattering SANS cross section for $T_{AEP}$-annealed Dy5. (b) d$\Sigma$/d$\Omega$ for Dy5 in different states and with an applied field of 0.05 T. (c) Normalized correlation function of Dy5 for different heat treatments. (d) Power-law exponent *n* as a function of heat-treatment temperature for total scattering d$\Sigma$/d$\Omega$ and for magnetic scattering d$\Sigma_M$/d$\Omega$. The *Q* range is restricted to 0.015 Å$^{-1}$ < *q* < 0.03 Å$^{-1}$.

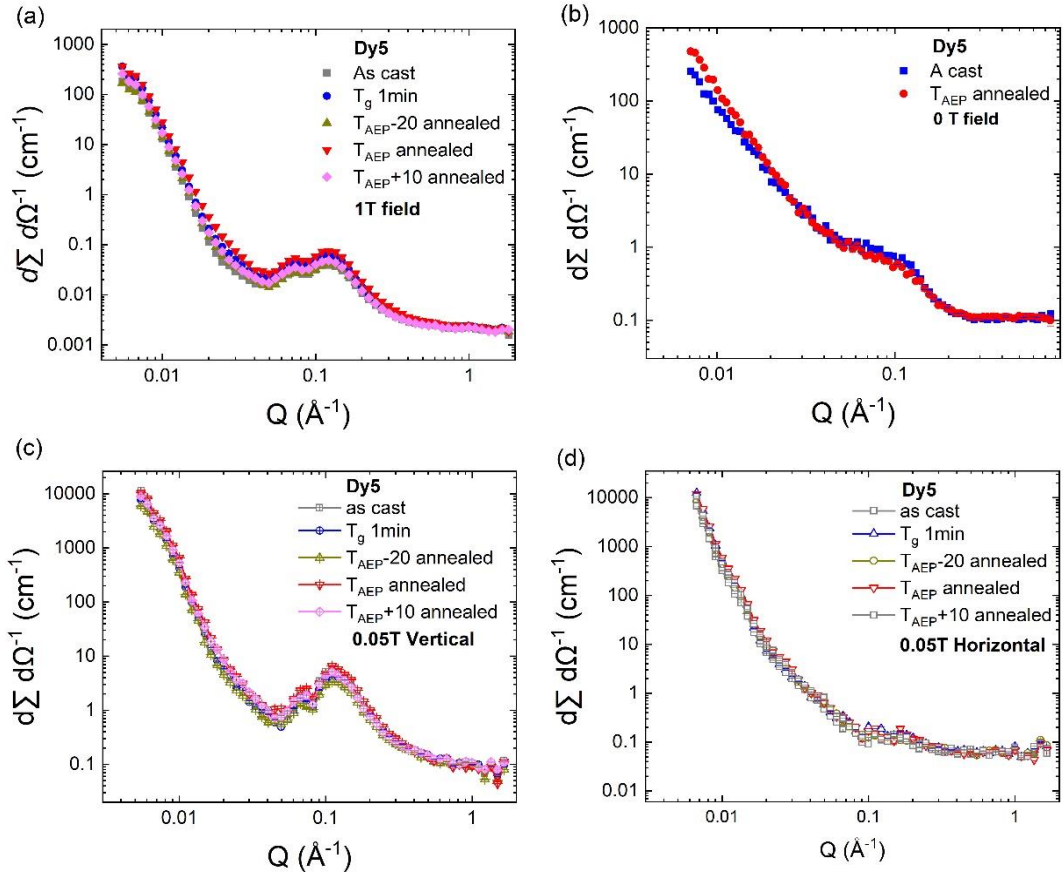

**Fig. 5.** (a) $d\Sigma/d\Omega$ for Dy5 in different states at 1 T field. (b) $d\Sigma/d\Omega$ for Dy5 in as-cast state and $T_{AEP}$-annealed state with no magnetic field. (c) Magnetic azimuthally averaged SANS cross sections in the vertical direction, and (d) magnetic azimuthally averaged SANS cross sections in the horizontal direction under an applied magnetic field of 0.05 T.

Fig. 6 shows a clearer view of the 2D SANS cross sections of Dy5 at different states by reducing $|Q_{max}|$ to 0.15 Å$^{-1}$. The samples as-cast, $T_{AEP}$ annealed, and crystal all show the elliptical geometry pattern under no magnetic field [Figs. 6(a), 6(c), 6(e)]. However, a hexagon-shaped geometry appears in all 2D scattering patterns under the 0.05 T field [Figs. 6(b), 6(d), 6(f)]. The spots with higher intensity appear at approximately ±0.12 Å$^{-1}$, which is consistent with the interference peaks in Fig. 4(b). Fig. 6(g) shows the scattering intensity as a function of azimuthal angle, and the six intensity maxima at intervals of about 60° appear under an applied magnetic field of 0.05 T. The disappearance of these characteristic scattering features under zero field suggests they are due to magnetic topological ordering. The hexagonal pattern of the $T_{AEP}$-annealed Dy5 BMG appears to be more ordered than that of the as-cast Dy5 BMG. The hexagonal pattern is further strengthened in the crystal sample, indicating a strong correlation between the magnetic and structural ordering.

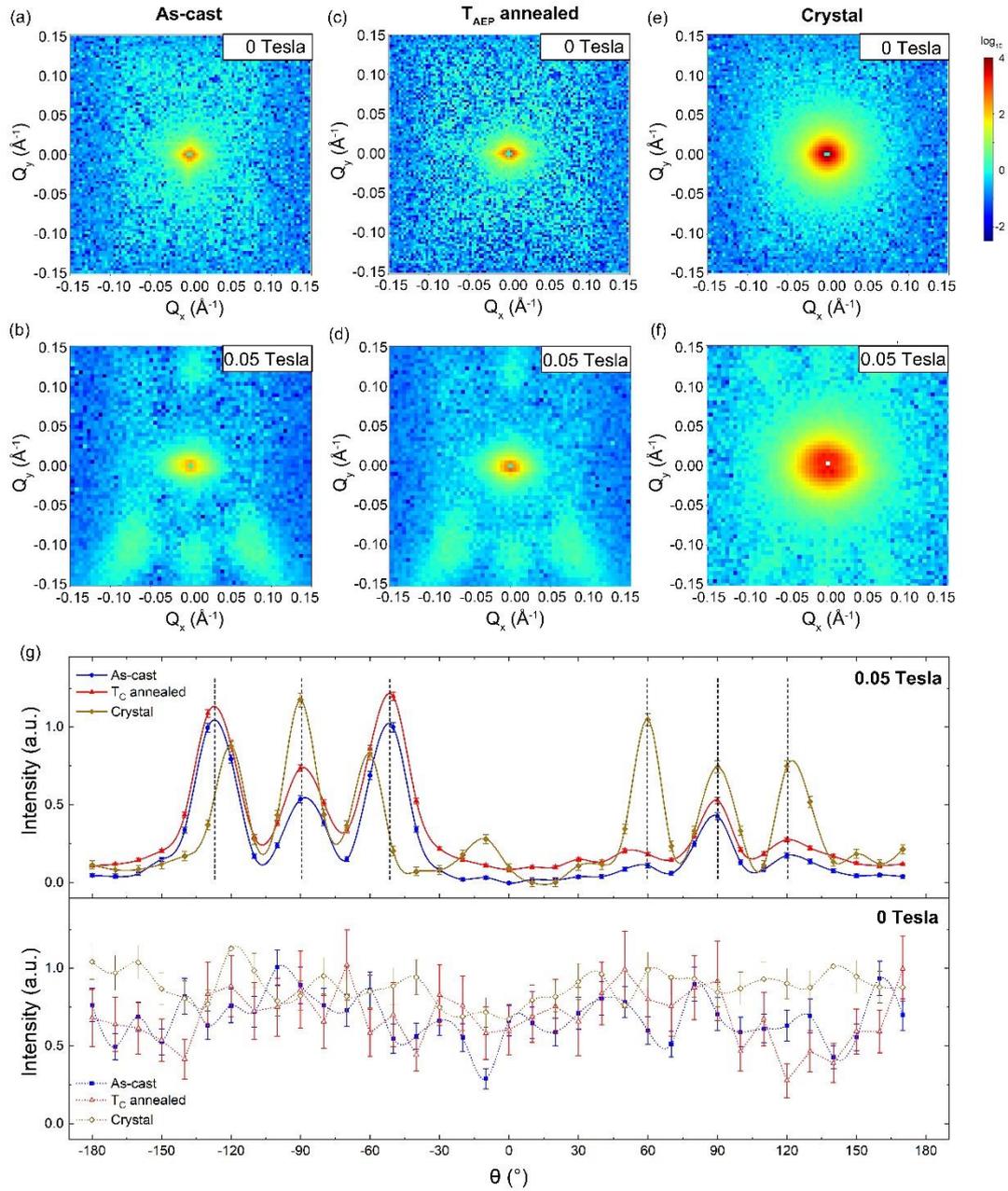

**Fig. 6** 2D total scattering SANS cross section for (a) as-cast, (c) $T_{AEP}$-annealed, and (e) crystal Dy5 under zero magnetic field with a reduced $Q$ range. 2D total scattering SANS cross section for (b) as-cast, (d) $T_{AEP}$-annealed, and (f) crystal Dy5 under 0.05 field (measured at CSNS). (g) Normalized intensity under 0.05 T (upper) and 0 T (lower) at different azimuthal angles from −180° to 180°.

In comparison, the scattering feature does not appear in compositions without the AEP [i.e., $(Fe_{0.73}(Dy_{0.5}Y_{0.5})_{0.03}B_{0.24})_{96}Nb_4$ (abbreviated Dy1.5Y1.5) and $(Fe_{0.73}Tm_{0.03}B_{0.24})_{96}Nb_4$ (abbreviated Tm3) MGs under the same field strength; see Fig. 7]. The critical casting size of $(Fe_{0.73}Dy_{0.03}B_{0.24})_{96}Nb_4$ is too small to satisfy for the requirements of SANS experiments due to the limited GFA. Here, we use them as substitutes for comparison. These two compositions are very similar, indicating that the magnetic topology correlates positively with the structural order at a medium-range scale [36, 37]. Fig. 7(a) shows the AEP is also exist in the supercooled-liquid region for $(Fe_{0.71}Tm_{0.05}B_{0.24})_{96}Nb_4$. Except for replacing Dy with Tm, the selected Fe-Tm-B-Nb MGs have the same composition, so they are assumed to have very similar properties. However, magnetic order exists in $(Fe_{0.71}Tm_{0.05}B_{0.24})_{96}Nb_4$ (abbreviated Tm5) but not in the $(Fe_{0.73}Tm_{0.03}B_{0.24})_{96}Nb_4$ (abbreviated Tm3), indicating that the magnetic topology may originate from the unique locally favored structures with topological order, which appear only in systems with AEPs.

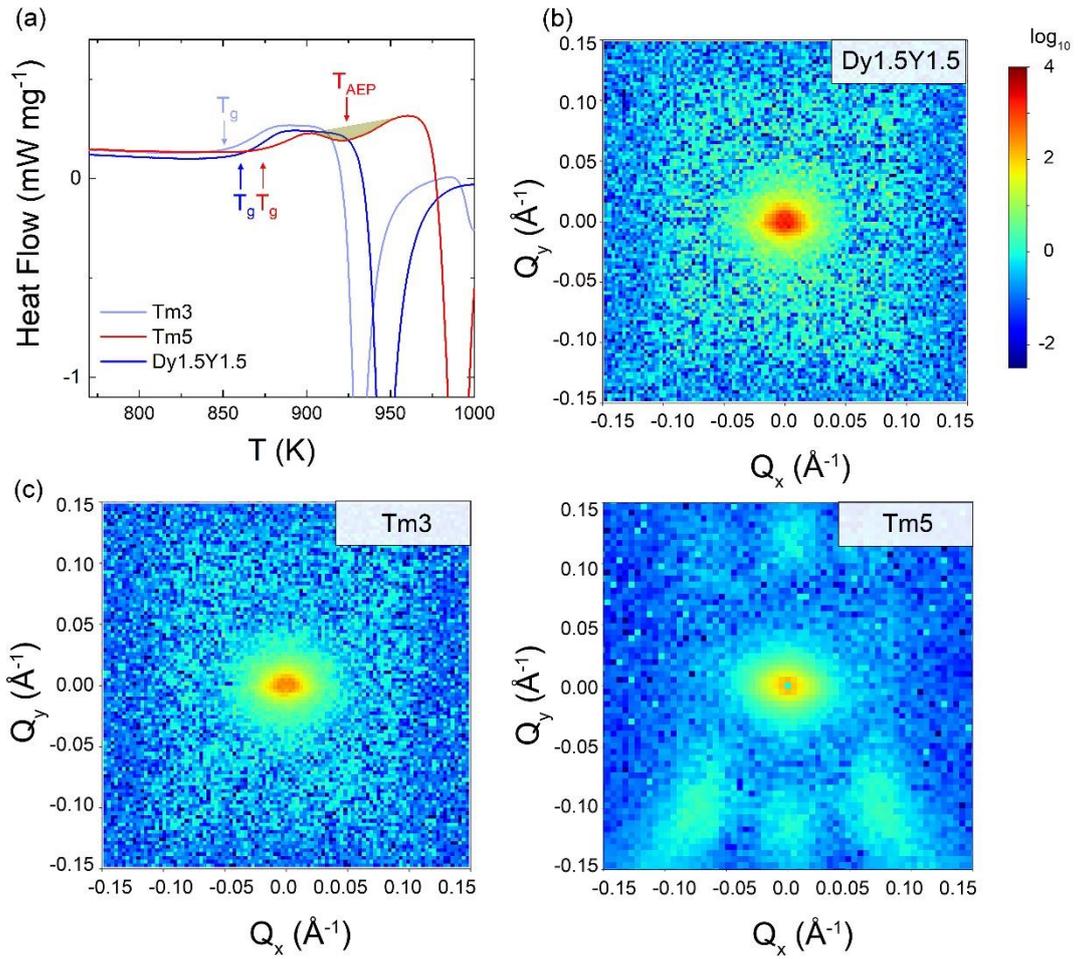

**Fig. 7** DSC curves for $(Fe_{0.73}(Dy_{0.5}Y_{0.5})_{0.03}B_{0.24})_{96}Nb_4$ (abbreviated Dy1.5Y1.5), $(Fe_{0.73}Tm_{0.03}B_{0.24})_{96}Nb_4$ (abbreviated Tm3) and $(Fe_{0.71}Tm_{0.05}B_{0.24})_{96}Nb_4$ (abbreviated Tm5) MGs. 2D total scattering cross section for (b) Dy1.5Y1.5 (c) Tm3 and (d) Tm5 with an applied magnetic field of 0.05 T.

## 4. Discussion

### 4.1 Medium-range ordering during LLPT

The *in-situ* synchrotron XRD suggests a structural ordering process dominated by MRO that occurs during the LLPT. The $S(Q)$ data first rule out the influence of crystallization, for no diffraction peaks from a crystal appear in $S(Q)$. Moreover, $Q_1$ and the peak-height evolution suggest an ordering process on the medium-range length scale. The heterogeneity length scale is at the MRO level [11, 38]. The PDF analysis further shows that cluster connectivity on the medium-range length scale plays a vital role in the LLPT, whereas the SRO changes little. This phenomenon is consistent with previous research in Fe-based BMGs with AEP [23]. Here we demonstrate that the edge-sharing cluster connection transformation dominates the evolution of MRO: the clusters tends to connect with each other via a higher-percentage edge-sharing scheme, forming a more ordered MRO structure.

Note that the slope changes around 0.12 Å$^{-1}$ in the SANS curves [Fig. 5(b)], indicating that a heterogeneity of about 20 Å is inherent in the amorphous Dy5 samples. Furthermore, a structure of the MRO with specific topological order was recently reported in the Pd-Ni-P BMG with a LLPT [38]. The MRO is constructed by six tricapped trigonal prism (TTPs) short-range motifs via an edge-sharing scheme. Given that the TTP cluster is also one of the primary clusters in Fe-based MGs [38, 39], a medium-range topological order may also exist, resulting in magnetic topological order. In addition, the hexagonal geometry in the 2D SANS pattern [39] only exists in compositions with AEPs (i.e., in Dy5 and Tm5 MGs). In comparison, Dy1.5Y1.5 and Tm3 do not have a defined topological geometry in spite of only a slight component bias. The rare-earth atoms play a glue-joint atom role on the medium-range scale because of their large atomic radius and the influence of enthalpy of mixing [8], therefore they may contribute to the formation of some MRO with topological order that triggers the LLPT.

*4.2 Correlation among medium-range order, magnetic nanodomains, and magnetic performance*

The SANS results suggest that a magnetic nanodomain exists with topological order and evolves with the transition to MRO. The magnetic performance is enhanced during the medium-range ordering process, including saturation magnetization and stress impedance. The corresponding magnetic correlation length $l_C$ is just at the medium-range scale, which changes with the evolution of the edge-sharing connection at medium-range scale during the LLPT [Fig. 8(a)]. As shown in Fig. 8(b), MRO ordering induces locally ordered inhomogeneous magnetization regions, which leads to interference peaks in the SANS curves. Moreover, the 2D SANS patterns indicate that locally ordered regions may possess hexagonal symmetry, giving rise to nanodomains with topological order. To eliminate the backscattering effect, we covered the backscattering detector with a Cd plate. Nevertheless, the interference peak persists at $Q \approx 0.12$ Å$^{-1}$, confirming that this scattering features results from the inherent magnetic structure. The decoding of the configuration of magnetic MRO needs further research.

Regarding the enhanced magnetic performances, note that ferromagnetism in amorphous materials is directly related to the exchange interaction between spins of neighboring unpaired electrons [40]. The PDF analysis indicates a reduced Fe–Fe bond after $T_{AEP}$ annealing [41] (Fig. S6), thus obtaining a stronger exchange interaction [1, 42]. Also, the enhancement of structural order also strengthens the local ferromagnetic order [43]. The above factors together promote the increase of saturation magnetization. For this study, the medium-range ordering induces the increase of local heterogeneity, which in turn enhances magnetic anisotropy[44]. As a result, during the magnetization process shown in Fig. 8(c), these regions will have more difficulty reaching saturation than the matrix, which is reflected in the increasing coercivity. Under saturation conditions, all magnetic moments are aligned in the same direction, which increases the overall magnetization (Fig. 8(d)). The stress-impedance effect is responsible for a change in the magnetic anisotropy that influences the permeability via an applied stress [3, 45, 46]. Here, the LLPT-induced magnetic anisotropy improves this effect.

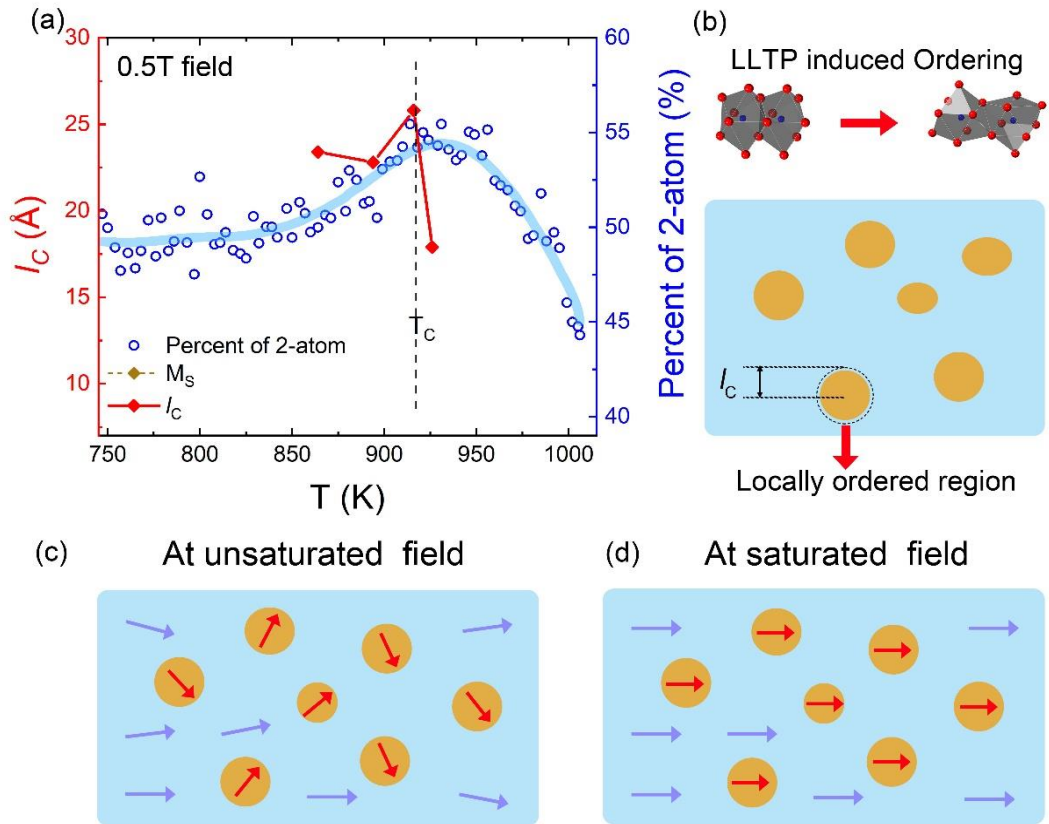

**Fig. 8.** (a) Evolution of 2-atom cluster connectivity, saturation magnetization, magnetic correlation length, and intraparticle correlation length as a function of temperature upon heating Dy5 BMG. (b)–(d) Response of nanoscale magnetic spin-misalignment structure during LLPT.

## 5. Conclusion

In summary, *in-situ* synchrotron high-energy x-ray diffraction and small-angle neutron scattering under a magnetic field suggest that the liquid-liquid phase transition of Fe-Dy-Nb-B BMG is dominated by a medium-range order transformation, which is manifested as a structural ordering with the enhancement of 2-atom connectivity. Nanodomains with topological order are found to exist in composition with the liquid-liquid phase transition and manifest as hexagonal patterns in small-angle neutron-scattering profiles. The Medium-range ordering can induce the nanodomains to be more locally ordered to form locally ordered regions, which have stronger exchange interactions due to the reduced Fe–Fe bond and structural order, which improves the saturation magnetization after the LLPT. Medium-range ordering also increases the local heterogeneity, which enhances magnetic anisotropy. Finally, the enhanced magnetic anisotropy facilitates the permeability response under applied stress, which improves the stress-impedance effect.

**Author contributions**

**Acknowledgments**

This study was financially supported by the National Key R&D Program of China (No. 2021YFB3802800), the Natural Science Foundation of Jiangsu Province (No. BK20200019) and the National Natural Science Foundation of China (Nos. 52222104, 12261160364, 51871120, and 51520105001). Si Lan acknowledges support from the Guangdong-Hong Kong-Macao Joint Laboratory for Neutron Scattering Science and Technology. X.-L. Wang acknowledges the support of the Shenzhen Science and Technology Innovation Commission (No. JCYJ20200109105618137). Yubin Ke acknowledges the support of the National Natural Science Foundation of China (No. 12275154) and the Guangdong Basic and Applied Basic Research Foundation (No. 2021B1515140028). This research used the resources of the China Spallation Neutron Source in Dongguan, China and the Advanced Photon Source, a US Department of Energy (DOE) Office of Science User Facility operated for the DOE Office of Science by Argonne National Laboratory under Contract No. DE-AC02-06CH11357, which was supported by the US DOE Office of Science, Office of Basic Energy Sciences. The small-angle neutron scattering experiments were carried out at the Japan Proton Accelerator Research Complex at the Tokai site of the Japan Atomic Energy Agency, JAEA in Ibaraki Prefecture, Japan and the Spallation Neutron Source.

**Data availability**

The data that support the findings of this study are available from the corresponding author upon reasonable request

**Declaration of Competing Interest**

The authors declare that they have no known competing financial interests or personal relationships that could appear to influence the work reported in this paper.

# Supplementary Material for "Evolution of medium-range order and its correlation with magnetic nanodomains in Fe-B-Nb-Dy bulk metallic glasses"

Jiacheng Ge,[a] Yao Gu,[a] Zhongzhen Yao,[a] Sinan Liu,[a] Huiqiang Ying,[a] Chenyu Lu,[b] Zhenduo Wu,[b,c] Yang Ren,[b] Jun-ichi Suzuki,[d] Zhenhua Xie,[e] Yubin Ke,[e,f,*] He Zhu,[a] Song Tang,[a] Xun-Li Wang,[b,g] and Si Lan[a,b,*]

[a]Herbert Gleiter Institute of Nanoscience, School of Materials Science and Engineering, Nanjing University of Science and Technology, 200 Xiaolingwei, Nanjing 210094, China

[b]Department of Physics, City University of Hong Kong, Hong Kong SAR, China

[c]Center for Neutron Scattering and Applied Physics, City University of Hong Kong Dongguan Research Institute, Dongguan 523000, China

[d]Guangdong-Hong Kong-Macao Joint Laboratory for Neutron Scattering Science and Technology, 1 Zhongziyuan Road, Dalang, Dongguan 523803, China

[e]Japan Proton Accelerator Research Complex, Japan Atomic Energy Agency, Tokai, Japan

[f]The China Spallation Neutron Source, Dongguang 523803, China,

[g]Center for Neutron Scattering, City University of Hong Kong Shenzhen Research Institute, Shenzhen 518057, China

**Table S1**

Table 1. Thermophysical parameters $T_g$, $T_x$, $T_l$, $T_{rg}$, $\gamma$, and critical thickness for Fe-B-Nb-Dy BMGs.

| Composition | $T_g$ (K) | $T_x$ (K) | $T_{AEP}$ (K) | $T_l$ (K) | $T_{rg}$ | $\gamma$ |
|---|---|---|---|---|---|---|
| $(Fe_{0.72}Dy_{0.03}B_{0.24})_{96}Nb_4$ | 847 | 925 |  | 1462 | 0.579 | 0.401 |
| $(Fe_{0.72}Dy_{0.05}B_{0.24})_{96}Nb_4$ | 864 | 975 | 914 | 1468 | 0.589 | 0.418 |

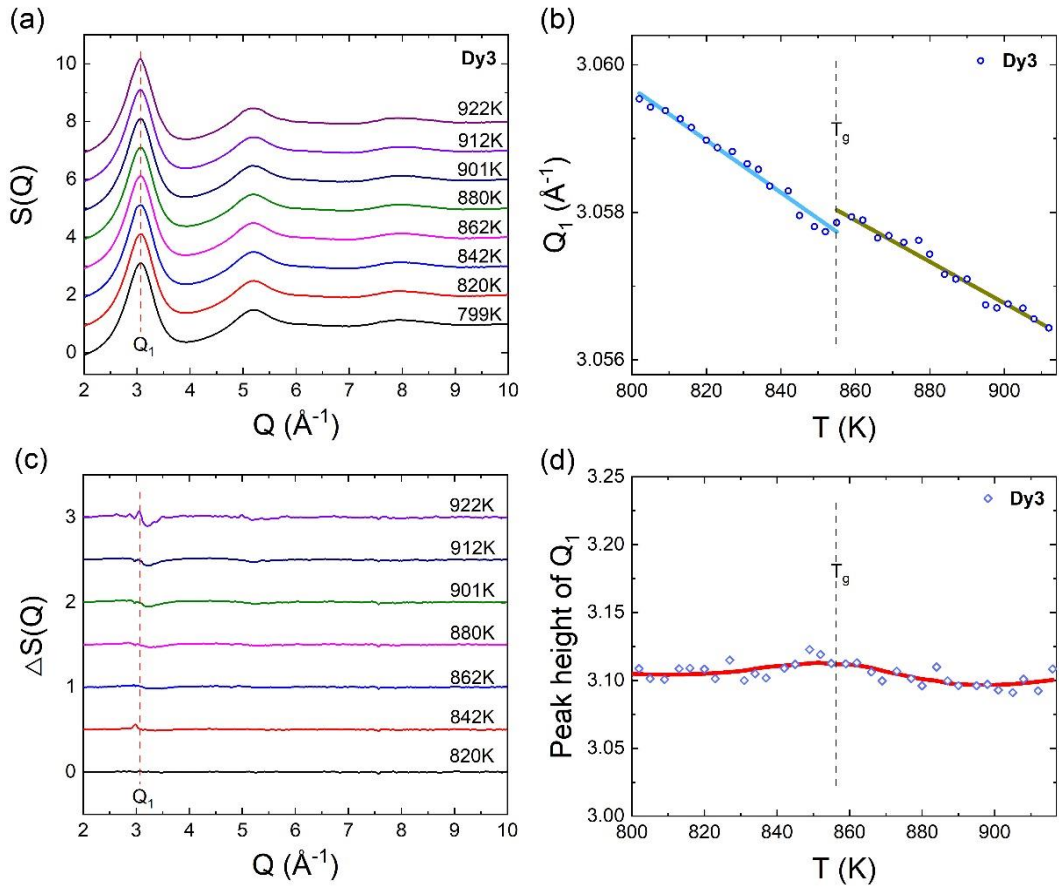

**Figure S1.** (a) $S(Q)$ for Dy3 MGs. (b) Evolution of peak position $Q_1$ for Dy3 BMG, showing only one change in slope at $T_g$. (c) Evolution of peak height of first diffraction peak for (c) Dy5. (d) Peak height as a function of temperature.

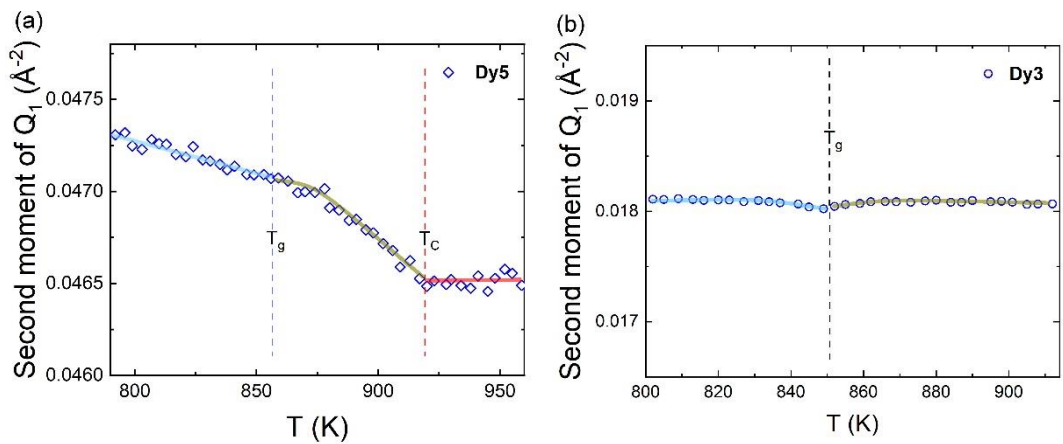

**Fig. S2** Evolution of second moment of $Q_1$ for (a) Dy5 and (b) Dy3 BMGs. The second moment of $Q_1$ is correlated with the structural order.

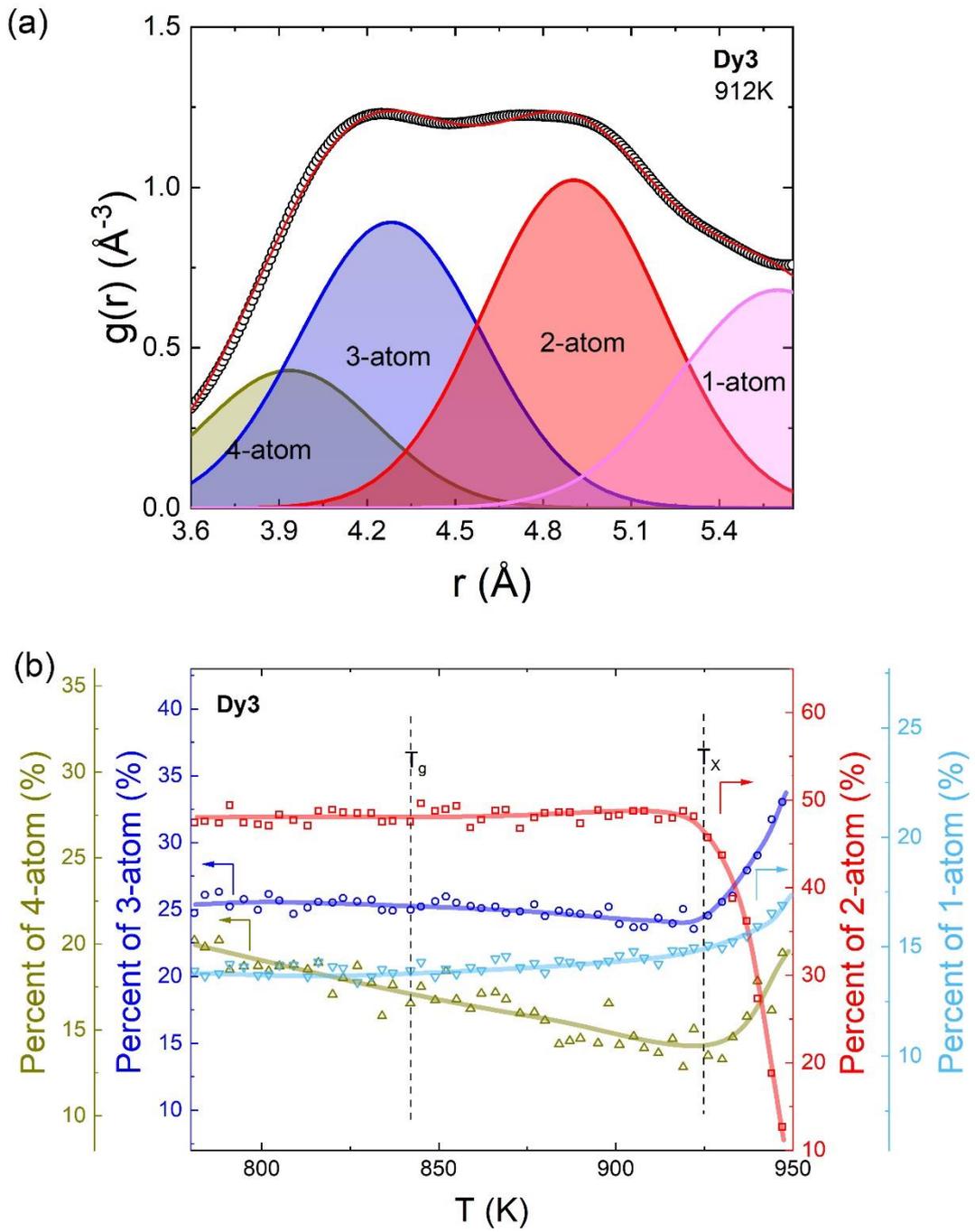

**Fig. S3** (a) *g*(*r*) upon heating Dy3 BMG and (b) evolution of the corresponding atomic connectivity, showing the percent of connection modes as a function of temperature.

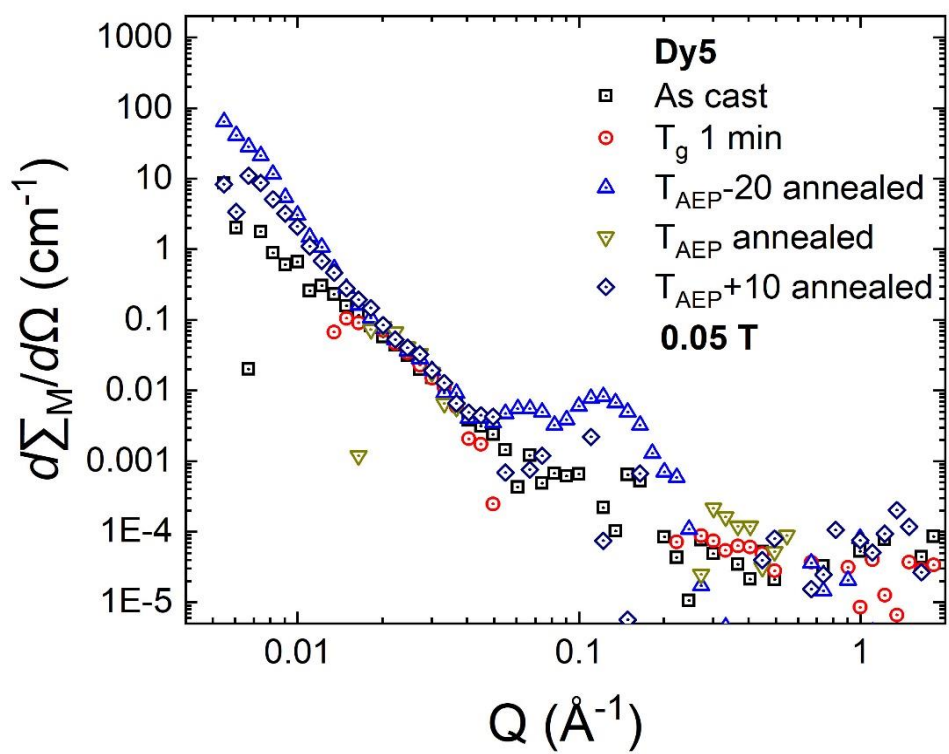

**Fig. S4** Magnetic spin-misalignment SANS cross sections under 0.05 T magnetic field, obtained by subtracting the scattering cross section $d\Sigma/d\Omega$ at a saturating magnetic field of 1 T from the scattering cross section at 0.05 T.

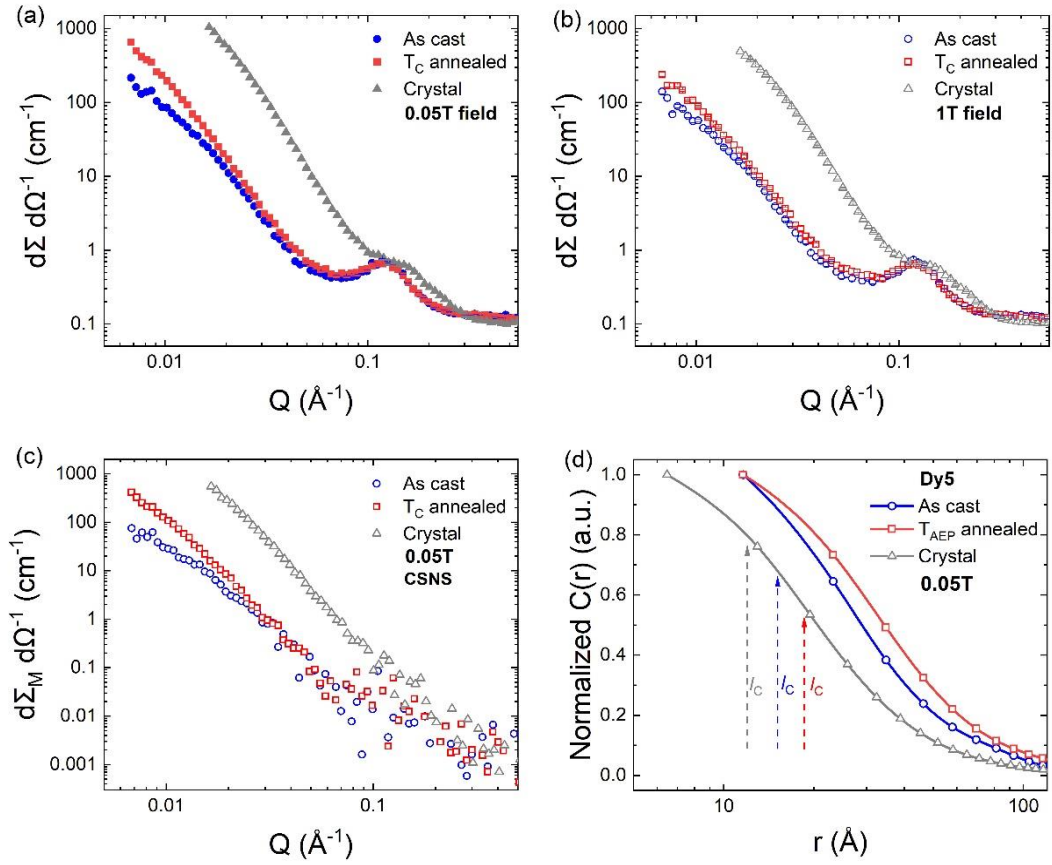

**Fig. S5** One-dimensional plot of total scattering cross section for as-cast, $T_{AEP}$-annealed, and crystal Dy5 at (a) 0.05 T and (b) 1 T field. (c) Corresponding spin-misalignment scattering cross section at 0.05 T. (d) Normalized correlation function for different states of Dy5.

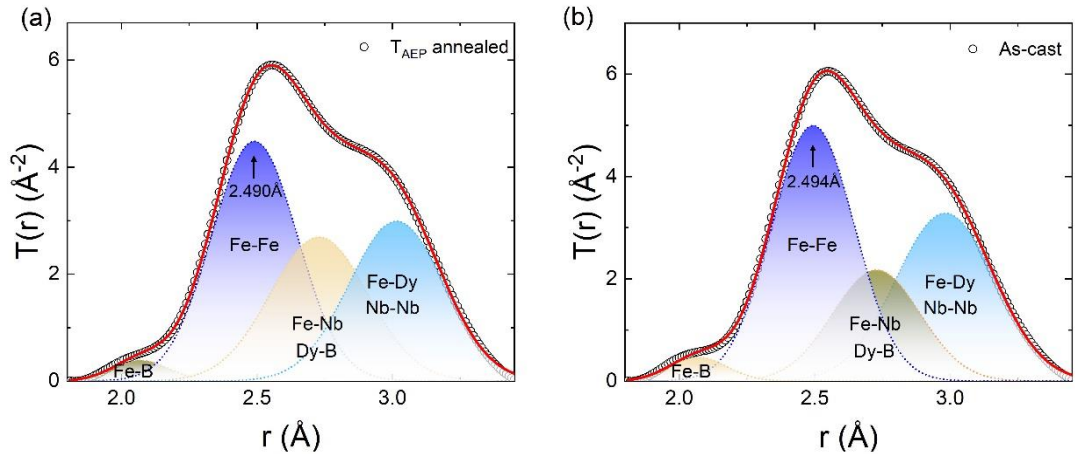

**Fig. S6** Fit of a linear combination of Gaussians to first-shell peaks of x-ray $T(r)$, obtained by $4\pi\rho_0 r + G(r)$, where $\rho_0$ is the number density. (a) Fit of $T_C$-annealed Dy5

and (b) fit of as-cast Dy5.